\documentclass{jetpl}
\twocolumn

\lat

\title{Spatial Kasner solution and an infinite slab with constant energy density}

\rtitle{Spatial Kasner solution and an infinite slab with constant energy density }

\sodtitle{Spatial Kasner solution and an infinite slab with constant energy density}

\author{Alexander\,Yu.\, Kamenshchik$^{+*}$\thanks{e-mail: kamenshchik@bo.infn.it},\ 
Tereza\, Vardanyan$^{\dagger}$}

\rauthor{A.\,Yu.\,Kamenshchik,  T. Vardanyan}

\sodauthor{Kamenshchik, Vardanyan}

\address{
$^+$ Dipartimento di Fisica e Astronomia, Universit\`a di Bologna and INFN, via Irnerio 46, 40126 Bologna, Italy\\
$^*$L.D. Landau Institute for Theoretical Physics RAS,
117940 Moscow, Russia\\
$^{\dagger}$Dipartimento di Fisica e Chimica, Universit\`a di L'Aquila, 67100 Coppito, L'Aquila, Italy\\ and 
INFN, Laboratori Nazionali del Gran Sasso, 67010 Assergi, L'Aquila, Italy }

\dates{}{}
\abstract{We  study the solutions of the Einstein equations in the  presence of a thick infinite slab with constant energy density. When there is an isotropy in the plane of the slab, we find an explicit exact solution that matches with the Rindler and Weyl-Levi-Civita spacetimes outside the slab. We also show that there are solutions that can be matched with general anisotropic Kasner spacetime outside the slab. In any case, it is impossible to avoid the presence of the Kasner type singularities in contrast to the well-known case of spherical symmetry, where by matching the internal Schwarzschild solution with the external one, the singularity in the center of coordinates can be eliminated.}
 
\begin{document}

\maketitle

The Kasner solution \cite{Kasner} of the Einstein equations for an empty universe having the spatial geometry of Bianchi-I type was one of the first exact solutions of General Relativity. Usually this solution is presented in the ``cosmological form'':
\begin{equation}
ds^2=dt^2-a_0^2t^{2p_1}dx^2-b_0^2t^{2p_2}dy^2-c_0^2t^{2p_3}dz^2.
\label{Kasner2}
\end{equation}
This form of the Kasner metric was rediscovered in papers \cite{Taub,Heck-Schuck,Khal-Lif} and has played an important 
role in cosmology. The study of Kasner dynamics in  paper \cite{Khal-Lif} has led to the discovery of the oscillatory approach to the cosmological singularity \cite{BKL}, known also as the Mixmaster universe \cite{Misner}. The further development of this line of research has brought the establishment of the connection between the chaotic behavior of the universe in superstring  models and the infinite-dimensional Lie algebras \cite{Damour}. 
 
However, in the original paper by Kasner \cite{Kasner} the positive definite metric with the dependence on one coordinate was considered. Introducing the normal spacetime signature, one can recover not only the cosmological metric (\ref{Kasner2}), but also a stationary metric that depends on one spatial coordinate:
 \begin{eqnarray}
 &&ds^2=a_0^2(x-x_0)^{2p_1}dt^2-dx^2-b_0^2(x-x_0)^{2p_2}dy^2\nonumber \\
 &&-c_0^2(x-x_0)^{2p_3}dz^2.
\label{Kasner3}
\end{eqnarray} 
The detailed story of different forms of Kasner metric is described in  Ref. \cite{Harvey}. 
Let us note that the metric (\ref{Kasner3}) has a singularity at the hypersurface $x=x_0$, where the value $x_0$ is quite arbitrary. On the other hand, in the cosmological Kasner solution (\ref{Kasner2}) the moment of time, when the universe is singular is traditionally chosen as $t=0$. For both solutions the Kasner indices $p_1,p_2$ and $p_3$ satisfy the relations
\begin{equation}
p_1+p_2+p_3=p_1^2+p_2^2+p_3^2 = 1.
\label{Kasner1}
\end{equation} 
 A convenient parametrization of the Kasner indices was presented in paper \cite{Khal-Lif}: 
\begin{eqnarray}
p_1=-\frac{u}{1+u+u^2},\ p_2 = \frac{1+u}{1+u+u^2},\ p_3 = \frac{u(1+u)}{1+u+u^2}.
\nonumber\\
\label{Lif-Khal}
\end{eqnarray}
It is interesting to compare the spatial Kasner solution (\ref{Kasner3}) with another famous static solution of the Einstein equations -- the  external spherically symmetric Schwarzschild solution \cite{Schwarz}. This solution has a singularity in the center of coordinates. To avoid it and to describe real spherically symmetric objects like stars, Schwarzschild also invented an internal solution \cite{Schwarz1} generated by a ball with constant energy density and isotropic pressure. At the boundary of the ball the pressure  disappears and the external and internal solutions are matched.   In this case there is no singularity in the center of the ball. Later, more general spherically symmetric geometries were studied in the papers by Tolman \cite{Tolman}, Oppenheimer-Volkoff \cite{Op-Vol}, Buchdahl  \cite{Buchdahl} and many others. Similar problems with cylindrical axial symmetry were also studied (see, e.g. \cite{Fulling0} and references therein). 

The solutions of the Einstein equations in the presence of an infinite plane or an infinite slab of a finite thickness with the metric 
\begin{equation}
ds^2=a^2(x)dt^2-dx^2-b^2(x)dy^2-c^2(x)dz^2
\label{plane}
\end{equation}
were also discussed in literature \cite{Amundsen,Fulling}. When $b(x)=c(x)$, these solutions are matched with special cases of the  Kasner metric (\ref{Kasner3}) such as the Rindler solution \cite{Rindler} with $p_1=1$, $p_2=p_3=0$ and the Weyl-Levi-Civita solution \cite{Weyl,Levi-Civita} with $p_1=-\frac13$, $p_2=p_3=\frac23$. 

In our paper \cite{we} we found two exact solutions in the spacetime with an infinite slab of thickness $2L$. In the first solution the pressure is isotropic, while in the second solution the tangential components of  pressure are equal to zero everywhere inside the slab. In both cases  pressure vanishes at the boundaries of the slab. Outside  the slab these solutions are matched with the Rindler spacetime and with the Weyl-Levi-Civita spacetime. 
In the present paper we describe general properties of the solutions of the Einstein equations when there is an isotropy in $yz$-plane, i.e. $b(x) = c(x)$, and explicitly construct a particular exact solution that differs from two solutions found in paper \cite{we}. Besides, we  discuss solutions with $b(x)\neq c(x)$, that are matched in the empty part of the space with the general Kasner solutions and not with its particular cases where $p_2=p_3$. We are not able to write down an explicit solution of this kind, however, analyzing the corresponding differential equations we can show that such solutions do exist.

For the metric (\ref{plane}) the Einstein equations inside the slab with constant energy density $\rho$ are
\begin{equation}
-\frac{b''}{b}-\frac{c''}{c}-\frac{b'c'}{bc}=\rho,
\label{Ein1}
\end{equation}
\begin{equation}
\frac{a'b'}{ab}+\frac{a'c'}{ac}+\frac{b'c'}{bc}=p_x,
\label{Ein2}
\end{equation}
\begin{equation}
+\frac{a''}{a}+\frac{c''}{c}+\frac{a'c'}{ac}=p_y,
\label{Ein3}
\end{equation}
\begin{equation}
+\frac{a''}{a}+\frac{b''}{b}+\frac{a'b'}{ab}=p_z,
\label{Ein4}
\end{equation}
where ``prime'' means the derivative with respect to $x$ and $p_x,p_y$ and $p_z$ are the corresponding components of pressure. Introducing new functions 
\begin{equation}
A = \frac{a'}{a},\ B=\frac{b'}{b},\ C = \frac{c'}{c},
\label{new}
\end{equation}
we can rewrite the Einstein equations (\ref{Ein1})-(\ref{Ein4}) as follows:
\begin{equation}
-B'-B^2-C'-C^2-BC = \rho,
\label{Ein1n}
\end{equation}
\begin{equation}
AB+AC+BC=p_x,
\label{Ein2n}
\end{equation}
\begin{equation}
A'+A^2+C'+C^2+AC = p_y,
\label{Ein3n}
\end{equation}
\begin{equation}
A'+A^2+B'+B^2+AB=p_z.
\label{Ein4n}
\end{equation}
 
We would like to find solutions of these equations inside the slab such that the pressure vanishes on its boundary. First of all, we should find the functions $B$ and $C$ that satisfy Eq. (\ref{Ein1n}). Then we should choose such a function $A$ that guarantees the disappearance of all three components of pressure at the boundary $x = \pm L$. This is impossible to do if the functions $B$ and $C$ do not satisfy some additional conditions. Thus, we should find such conditions explicitly and choose solutions of Eq. (\ref{Ein1n}) that satisfy these conditions. Then we should choose a function $A$ that provides the disappearance of the pressure on the boundary. After that one can match an obtained solution with the external solutions in the empty half-spaces $x > L$ and $x < -L$.

If there is an isotropy in the plane of the slab ($yz$-plane), then $B=C$ and Eq. (\ref{Ein1n}) has the general solution \cite{we}:
 \begin{equation}
 B=C = -\frac23 k \tan k(x+x_0),
 \label{Sym}
 \end{equation}
where 
 \begin{equation}
 k = \frac{\sqrt{3\rho}}{2}.
 \label{k}
 \end{equation}
Let us suppose now that at one of the boundaries, say $x=-L$, one has $B(-L)=C(-L)=0$. Then Eq. (\ref{Ein2n}) does not impose restrictions on the value of $A(-L)$, while from Eq. (\ref{Ein3n}) it follows that 
\begin{equation}
A'(-L)+A^2(-L)=-C'(-L). 
\label{condition}
\end{equation}
It follows immediately from Eq. (\ref{Sym}) that if $C(-L) = 0$, then $C'(-L) = -\frac23 k^2$. Hence,
\begin{equation}
A'(-L)+A^2(-L)=\frac23 k^2. 
\label{condition1}
\end{equation}
To have $C(-L) = 0$, one should choose $x_0=L$. Then at $x=L$ we shall have 
\begin{eqnarray}
&&B(L)=C(L)=-\frac23k\tan 2kL,\ A(L) = \frac13k\tan 2kL, \nonumber \\
&& B'(L)=C'(L) = -\frac23\frac{k^2}{\cos^2 2kL}, \nonumber \\
&& A'(L)=-\frac13k^2\tan^2 2kL+\frac23\frac{k^2}{\cos^2 2kL}. 
\label{condition2}
\end{eqnarray}

Now we can choose an arbitrary function $A(x)$ satisfying the boundary conditions (\ref{condition1})-(\ref{condition2}) and in this case all the components of the pressure disappear at the boundary of the slab. The simplest function that satisfies these three boundary conditions is a quadratic function
\begin{equation}
A(x) = \alpha(x-L)^2+\beta(x-L)+\gamma.
\label{A-quadr}
\end{equation}
Equations (\ref{condition2}) give immediately 
\begin{eqnarray}
&&\beta=A'(L)=-\frac13k^2\tan^2 2kL + \frac23\frac{k^2}{\cos^2 2kL},\nonumber \\
&&\gamma = A(L)=\frac13 k \tan 2kL.
\label{A-quadr1}
\end{eqnarray}
Eq. (\ref{condition1}) instead gives a quadratic equation for the coefficient $\alpha$:
\begin{eqnarray}
&&16\alpha^2L^4+\alpha(-4L-16\beta L^3+8\gamma L^2)\nonumber \\
&&+\beta+4\beta^2L^2+\gamma^2-4\beta\gamma L-\frac23k^2 = 0.
\label{A-quadr2}
\end{eqnarray} 
Substituting the values of $\beta$ and $\gamma$ from Eq. (\ref{A-quadr1}) into Eq. (\ref{A-quadr2}), we obtain the following solutions:
\begin{eqnarray}
&&\alpha_{1,2} = \frac{1}{8L^3}\left[\frac{11}{12}+\frac83k^2L^2+\frac{1}{12}(4kL\tan 2kL-1)^2\right. \nonumber \\
&&\left. \pm \sqrt{\frac23+\frac{16}{3}k^2L^2+\frac13(2kL\tan 2kL-1)^2}\right].
\label{A-quadr3}
\end{eqnarray}
Note that the expression under the sign of the square root is positive and these two solutions for the coefficient $\alpha$ always exist. Thus, substituting the expressions (\ref{A-quadr3}) and (\ref{A-quadr1}) into Eq. (\ref{A-quadr}) we obtain two solutions for the function $A(x)$, which together with the expression (\ref{Sym}) with $x_0 = L$ provide the implementation of the boundary condition for all components of the pressure. Using Eqs. (\ref{Ein2n}) and (\ref{Ein3n}) we can write down the explicit expressions for the tangential and perpendicular components of the pressure. 

The form of the functions $B(x) = C(x)$ is the same as in our preceding paper \cite{we}. Thus, the matching conditions at the boundaries $x=\pm L$ will give the same result. {For} $x > L$ we shall have a Weyl-Levi-Civita spacetime, while {for} $x < -L$ we shall have a Rindler spacetime. Let us emphasize that the solution described above differs from two exact solutions presented in our preceding paper \cite{we}.

Suppose now that $B(x)\neq C(x)$, and their values at the boundary $x=-L$ are also different
\begin{equation}
B(-L)=B_0,\ C(-L) = C_0.
\label{ani}
\end{equation}
Then, from the requirement of the disappearance of the normal component of the pressure   $p_x$ at the boundary and from Eq. (\ref{Ein2n}), it follows that 
\begin{equation}
A(-L)=A_0 = -\frac{B_0C_0}{B_0+C_0}.
\label{ani1}
\end{equation}
It is easy to see that these three numbers constitute a Kasner triplet, multiplied by $D/p_2$, namely
\begin{eqnarray}
&&A_0 = p_1\frac{D}{p_2},\nonumber \\
&&B_0 = D,\nonumber \\
&&C_0 = p_3\frac{D}{p_2},
\label{ani-Kas}
\end{eqnarray}
where $p_1, p_2$ and $p_3$ satisfy the relations (\ref{Kasner1}) and $D$ is an arbitrary constant. 
These Kasner indices are determined  by the relation between $B_0$ and $C_0$. Indeed, one can see that 
\begin{equation}
\frac{C_0}{B_0} = \frac{p_3}{p_2}=u,
\label{K-L}
\end{equation} 
where $u$ is the parameter introduced in \cite{Khal-Lif} and connected with the Kasner indices by the relations (\ref{Lif-Khal}). Let us look now at the equations (\ref{Ein3n}) and (\ref{Ein4n}). If we require the disappearance of the tangential components of the pressure $p_y$ and $p_z$ at $x=-L$, then we have the following relation connecting the values of the first derivatives $B'(-L)$ and $C'(-L)$:
\begin{eqnarray}
B'(-L)-C'(-L) = \frac{C^3(-L)-B^3(-L)}{B(-L)+C(-L)}=B_0^2\frac{u^3-1}{u+1}.
\label{deriv}\nonumber\\
\end{eqnarray}
 On the other hand, from Eq. (\ref{Ein1n}) we find 
 \begin{equation}
 B'(-L)+C'(-L) = -\rho-(1+u+u^2)B_0^2. 
 \label{deriv1}
 \end{equation}
 
Now it is convenient to  introduce a pair of two functions:
 \begin{eqnarray}
 &&F(x)\equiv B(x)+C(x),\nonumber \\
 &&G(x) \equiv B(x)-C(x).
 \end{eqnarray}
 In this case Eq. (\ref{Ein1n}) becomes
 \begin{equation}
 F'(x) + \frac34F^2(x) +\frac14G^2(x) = -\rho.
 \label{Ein1nn}
 \end{equation}  
Two functions $F(x)$ and $G(x)$ and the first derivative of one of them $F'(x)$ enter this equation. These functions satisfy the initial conditions
\begin{eqnarray}
 &&F(-L) = B(-L)+C(-L) = B_0(1+u),\nonumber \\ 
 &&G(-L) = B(-L)-C(-L) = B_0(1-u). 
 \label{initial}
 \end{eqnarray}  
The initial value for $F'(-L)$ is defined from Eq. (\ref{Ein1nn}). Besides, we have an additional initial condition on $G'(-L)$ (see Eq. (\ref{deriv1})) coming from the requirement of the simultaneous disappearance of all components of the pressure at the boundary $x=-L$. 

Now we would like to  match our internal solution, satisfying the boundary conditions described above, with the Kasner solution in the empty half-space $x \leq -L$. As was explained in our preceding paper \cite{we}, for this matching it is necessary that the functions $A, B$ and $C$ at the boundaries are continuous. This means that the values of the Kasner indices in the solution describing the empty half-space $x \leq -L$ coincide with those coming from Eqs. (\ref{ani-Kas})-(\ref{K-L}).  The derivative of the second scale factor of the empty solution (\ref{Kasner3}) at $x = -L$ should be equal to the value of $B_0$:
\begin{equation}
\frac{p_2}{-L-x_L} = B_0.
\label{match}
\end{equation}
Here $x_L$ indicates the location of the singularity in the Kasner solution. However, our solution (\ref{Kasner3}) is valid only for $x \leq -L$. Thus, if $x_L > -L$, the singularity will be absent. The requirement $x_L > -L$ implies the negativity of the left-hand side of the equality (\ref{match}) and, hence, its right-hand side also should be negative, i.e. $B_0 < 0$. Then it follows from Eq. (\ref{ani-Kas})   that $C_0 < 0$ while $A_0 > 0$. From the negativity of $B_0$ and $C_0$ follows the negativity of $F(-L)$. Then the Eq. (\ref{Ein1nn}) implies that the derivative $F'(x)$ is negative, moreover it satisfies an inequality 
\begin{equation}
F'(x) < -\rho - B_0^2(1+u)^2.
\label{ineq}
\end{equation}  
{This} means that at the other boundary $x=L$, the values $B(L)$ and $C(L)$ will be negative. Let us represent an empty space Kasner solution for $x\geq L$ as 
 \begin{eqnarray}
 &&ds^2=\tilde{a}_0^2(x-x_R)^{2\tilde{p}_1}dt^2-dx^2-\tilde{b}_0^2(x-x_R)^{2\tilde{p}_2}dy^2\nonumber \\
 &&-\tilde{c}_0^2(x-x_R)^{2\tilde{p}_3}dz^2,
\label{Kasner4}
\end{eqnarray}
with the singularity at $x=x_R$, and a triplet of the Kasner indices $\tilde{p}_1,\tilde{p}_2,\tilde{p_3}$. At the boundary we have the following condition 
\begin{equation}
\frac{\tilde{p}_2}{L-x_R} = B(L) < 0.
\label{ineq1}
\end{equation}
It follows from the relation (\ref{ineq1}) that $x_R > L$, i.e. that the singularity is present in the region of the validity of our solution. Thus, we have seen that it is impossible to construct such a solution in the slab that is matched with two non-singular Kasner solutions in two empty half-spaces outside the slab. 
  
Let us discuss in detail what happens with the solution of the Einstein equations in the slab that are matched with the Kasner solution outside the slab. As was said above, we have chosen some values of the functions $B$ and $C$ at the boundary $x=-L$. The requirement of the disappearance of the normal component of the pressure $p_x$ has given us the value of the function $A$ at the same boundary. The Einstein equation (\ref{Ein1n}) has given the value of $B'(-L)+C'(-L)$. Then the requirement of disappearance of tangential components of the pressure $p_y$ and $p_z$ at this boundary implies the condition  (\ref{deriv}) on the difference of the derivatives of $B$ and $C$, which permits us to find the value of $A'(-L)$ from Eq. (\ref{Ein4n}):
\begin{equation}
A'(-L) = \frac{\rho}{2} +\frac{u(1+u+u^2)B_0^2}{(u+1)^2}.
\label{A-prime}
\end{equation}
Let us suppose that we would like  to match the solution of the Einstein equations in the slab, satisfying the boundary conditions at $x=-L$ represented above, with some Kasner solution at $x=L$. Note that we cannot prescribe the values $B(L)$ and $C(L)$ in an arbitrary way: their sum should satisfy the inequality 
\begin{equation}
B(L) + C(L) = F(L) < 2L(-\rho-F^2(-L)). 
\label{ineq2}
\end{equation}
This inequality can be rewritten as 
\begin{equation}
\tilde{B}_0(1+\tilde{u}) < 2L(-\rho-F^2(-L)),
\label{ineq3}
\end{equation}   
where the parameter $\tilde{u}$ defines a new triplet of the Kasner indices.   It is easy to see that we can satisfy this condition for any value of $\tilde{u}$ 
choosing a  negative constant $\tilde{B}_0$, obeying this inequality. Then, knowing the values of the constants $B(L)$ and $C(L)$, we can find the value of $A(L)$ using the condition of the disappearance of the normal component of the pressure $p_x$ at the boundary $x=L$. Finally, to provide the disappearance of the tangential components of the pressure at this boundary, the derivative of the function $G(x) = B(x)-C(x)$ should satisfy the condition that has the same form  as the equality (\ref{deriv}). 

Let us analyse now the equation  (\ref{Ein1nn}). The function $G(x)$   should be chosen in such a way to  satisfy  the four boundary conditions on $G(-L), G'(-L), G(L)$ and $G'(L)$.   Then we obtain the first order differential equation for the function $F(x)$. Having defined the initial condition $F(-L)$, we have a well defined evolution of this function 
in the interval $x \in [-L,L]$. This evolution defines uniquely the value of the function $F$ at the boundary $x=L$. However, we would like to have some chosen value of $F(L)$. Can we modify our function $G(x)$ in such a way as to achieve this goal? It is surely possible. We can change our free function $G(x)$ without changing the four boundary conditions mentioned above.  If we change it in such a way that its absolute value increases in the open interval between $-L$ and $L$, then the derivative $F'(x)$ decreases and the final value of $F(L)$ also decreases. If, instead, the absolute value of $G(x)$ decreases, the values of $F'(x)$ and, hence, of $F(L)$ increase. Thus, there exist solutions $F(x)$ and $G(x)$ in the slab satisfying the chosen boundary conditions and smoothly matching with the empty space Kasner solutions for $|x| > L$. However, at least one of these Kasner solutions possesses a singularity.  

In conclusion,  we can say that studying such classical topic as the solutions of the Einstein equations possessing Bianchi type symmetries (see e.g. \cite{Bel,lectures}), for example,  the simplest Bianchi-I symmetry considered here or more complicated Bianchi-VIII symmetry \cite{Kirillov}, can still give interesting and unexpected results.

The authors are grateful to G. Venturi, R. Casadio and K. A. Bronnikov for useful discussions. The work of A.K. was partially supported by the RFBR grant 18-52-45016.


\vspace{2cm}
\begin{thebibliography}{99}
\bibitem{Kasner}
E.~Kasner,
  Am.\ J.\ Math.\  {\bf 43}, 217 (1921).
\bibitem{Taub}
A.~H.~Taub,
  Annals Math.\  {\bf 53}, 472 (1951).
\bibitem{Heck-Schuck}
O. Heckmann and E. Schucking, Handbuch der Physik {\bf 53}, 489 (1959).
\bibitem{Khal-Lif}
E.~M.~Lifshitz and I.~M.~Khalatnikov,
  Adv.\ Phys.\  {\bf 12}, 185 (1963).
\bibitem{BKL}
V.~A.~Belinsky, I.~M.~Khalatnikov and E.~M.~Lifshitz,
  Adv.\ Phys.\  {\bf 19}, (1970) 525.
\bibitem{Misner}
C.~W.~Misner,
  Phys.\ Rev.\ Lett.\  {\bf 22}, 1071 (1969).
\bibitem{Damour}
T.~Damour, M.~Henneaux and H.~Nicolai,
  Class.\ Quant.\ Grav.\  {\bf 20}, R145 (2003).
\bibitem{Harvey}
A. Harvey, Gen. Rel. Grav. {\bf 22}, 1433 (1990). 
\bibitem{Schwarz}
K.~Schwarzschild,
  Sitzungsber.\ Preuss.\ Akad.\ Wiss.\ Berlin (Math.\ Phys.\ ) {\bf 1916}, 189 (1916).
\bibitem{Schwarz1}
K.~Schwarzschild,
  Sitzungsber.\ Preuss.\ Akad.\ Wiss.\ Berlin (Math.\ Phys.\ ) {\bf 1916}, 424 (1916).
\bibitem{Tolman}
R.~C.~Tolman,
  Phys.\ Rev.\  {\bf 55}, 364 (1939).
\bibitem{Op-Vol}
J.~R.~Oppenheimer and G.~M.~Volkoff,
  Phys.\ Rev.\  {\bf 55}, 374  (1939).
\bibitem{Buchdahl}
H.~A.~Buchdahl,
  Phys.\ Rev.\  {\bf 116}, 1027 (1959).
\bibitem{Fulling0}
C.~S.~Trendafilova and S.~A.~Fulling,
  Eur.\ J.\ Phys.\  {\bf 32},  1663 (2011).
\bibitem{Amundsen}
P.~A.~Amundsen and O.~Gron,
  Phys.\ Rev.\ D {\bf 27}, 1731 (1983).
\bibitem{Fulling}
S.~A.~Fulling, J.~D.~Bouas and H.~B.~Carter,
  Phys.\ Scripta {\bf 90}, no.8,  088006  (2015).
\bibitem{Rindler}
W.~Rindler,
  Am.\ J.\ Phys.\  {\bf 34}, 1174 (1966).
\bibitem{Weyl}
H. Weyl, Annalen Physik {\bf 54}, 117 (1917).
\bibitem{Levi-Civita}
T. Levi-Civita, Atti Accad. Naz. Rend. {\bf 27}, 240 (1918).
\bibitem{we}
A.Yu. Kamenshchik and T. Vardanyan, 
Phys. Lett. B {\bf 792}, 430  (2019).
\bibitem{Bel}
V. Belinski and M. Henneaux, {\it The Cosmological Singularity}, (Cambridge University Press, Cambridge, 2017). 
\bibitem{lectures}
A. Yu. Kamenshchik, 
The Bianchi Classification of the Three-Dimensional Lie Algebras and Homogeneous Cosmologies and the Mixmaster Universe,
in {\it Einstein Equations: Physical and Mathematical Aspects of General Relativity. DOMOSCHOOL 2018}, edited by S. Cacciatori, B. G\"uneysu and S. Pigola,
(Birkh\"auser, Cham, 2019), pp. 93-137.
\bibitem{Kirillov}
A.~A.~Kirillov, G.~Montani and E.~P.~Savelova,
  JETP Lett.\  {\bf 107}, no. 6, 333 (2018)
  [Zh.\ Eksp.\ Teor.\ Fiz.\  {\bf 107}, no. 6, 349 (2018)].




\end{thebibliography}
\end{document}